\documentclass[twocolumn,showpacs,preprintnumbers,amsmath,amssymb]{revtex4}

\usepackage{graphicx}
\usepackage{dcolumn}
\usepackage{bm}

\begin{document}

\preprint{}

\title{Spin Susceptibility of a 2D Electron System in GaAs towards the Weak Interaction Region
}

\author{Y.-W. Tan$^{1}$}
\author{J. Zhu$^{1}$}
\author{H. L. Stormer$^{1,2,3}$}
\author{L. N. Pfeiffer$^{3}$}
\author{K. W. Baldwin$^{3}$}
\author{K. W. West$^{3}$}

\affiliation{$^{1}$Department of Physics, Columbia University, New
York, New York 10027\\
$^{2}$Department of Applied Physics and Applied Mathematics,
Columbia University, New York, New York 10027\\ $^{3}$Bell Labs,
Lucent Technologies, Murray Hill, New Jersey 07974
}%

\date{\today}

\begin{abstract}
We determine the spin susceptibility $\chi$ in the weak
interaction regime of a tunable, high quality, two-dimensional
electron system in a GaAs/AlGaAs heterostructure. The band
structure effects, modifying mass and g-factor, are carefully
taken into accounts since they become appreciable for the large
electron densities of the weak interaction regime. When properly
normalized, $\chi$ decreases monotonically from 3 to 1.1 with
increasing density over our experimental range from $0.1$ to
$4\times10^{11} cm^{-2}$. In the high density limit, $\chi$ tends
correctly towards $\chi\rightarrow 1$ and compare well with recent
theory.
\end{abstract}

\maketitle

\section{\label{sec:level1}Introduction}

The recent interests in a possible correlated many particle state
\cite{Bloch, Stoner, Wigner} at very low electron density of a two
dimensional electron system (2DES) have sparked a series of
experimental investigations in different materials.\cite{Shashkin,
Pudalov, Tutuc, Zhu, Vakili} In these experiments, the spin
susceptibility, $\chi$, which is proportional to the product of
effective mass, m*, and g-factor, g*, usually deviates from its
noninteracting ``bare" band edge value, $m_{0}g_{0}$, and
increases drastically with decreasing electron density. The focus
of these experiments is on the low density region, whereas $\chi$
has not been pursued to high densities. While we expect m*g* to
approach $m_{0}g_{0}$ in the high density limit, the explicit
behavior in the intermediate regime remains experimentally
unresolved.

The density dependence of $\chi$ is usually expressed in terms of
the dimensionless parameter, $r_{s}$. It is defined as the average
e-e distance measured in units of the Bohr radius, which is also
the ratio of Coulomb energy to Fermi energy. In 2D,
$r_{s}\propto1/ \sqrt[]{n}$, the intermediate density regime
resides at $r_{s}\sim1$ and the high density regime at
$r_{s}\rightarrow 0$. High $r_{s}$ measurements of $m^{*}g^{*}$
from different materials generally differ from each other and from
theory. Most early calculations\cite{Attaccalite, Tanatar,
Reimann, YG} produced the sharp rise of $\chi$ at low densities,
but differed quantitatively from experiments. Of those, data on
very thin AlAs\cite{Vakili} quantum wells came closest to theory,
probably since the latter omitted finite thickness effects.
Indeed, recent calculation \cite{dePalo, Zhang, Dharma} have shown
that most of the discrepancies arise from the finite thickness of
the 2DES and the resulting modification of the Coulomb
interaction. These calculations for $\chi$ extend far into the
weak interaction regime, yet there are presently no data available
to test their validity. An additional incentive for such
measurement is the fact that an empirical relation for $\chi$,
observed in the low-density regime \cite{Zhu}, when extrapolated
to high density would imply vanishing $\chi$, whereas one expects
$\chi(n\rightarrow\infty)=1$.

In this paper we provide high quality data on $\chi$ through the
intermediate interaction regime and towards small $r_{s}$ for
comparison with recent, realistic theoretical calculations. Once
the data are corrected for the energy dependence of band mass and
g-factor, theory and experiment reach a remarkable level of accord
and the data show the correct limiting behavior for large
densities.

\section{\label{sec:level1}Device structure and the tilted-field measuring techniques}

Our primary sample is a heterojunction-insulated gate field effect
transistor (HIGFET). It consists of a $5\mu m$ buffer GaAs layer
grown by molecular beam epitaxy (MBE) on a (001) GaAs substrate. A
subsequently grown, 200 period superlattice of 10nm
$Al_{0.33}Ga_{0.67}As$ and 3nm GaAs acts as a diffusion barrier.
It is followed by $5\mu m$ of undoped GaAs, which functions as the
channel. This layer is covered by 5nm of AlAs and $4\mu m$ of
$Al_{0.33}Ga_{0.67}As$, being the dielectric material, topped by
25nm of heavily doped GaAs, which serves as the gate. The specimen
is processed into a $600\mu m$ square mesa. We use standard
photolithography to define fifteen Ni-Ge-Au contact pads. Our
specimen has mobilities as high as $1\times10^{7}cm^{2}/Vs$. The
density, $n$, can be tuned by varing the gate voltage, $V_{g}$.
For this sample, $n(10^{10}cm^{-2})=0.15+16\times V_{g}(V)$. Such
a HIGFET allows us to measure a wide range of densities in a
single specimen, eliminating sample to sample parameter
fluctuations. For our sample, which has a density range spanning
from $0.1\times 10^{11}$ to $4\times 10^{11} cm^{-2}$, the $r_{s}$
value can be tuned from 0.9 to 6, covering a wide range of e-e
interaction strength.

All measurements were performed in a dilution refrigerator with a
base temperature of $\sim 30mK$. The sample was mounted on a
rotating sample holder to control separately the total magnetic
field, $B_{tot}$, and the field, $B_{\bot}$, perpendicular to the
2DES. We used conventional low frequency ac lock-in techniques
with excitation currents ranging from 1nA to 100nA, chosen to
avoid sample heating.

We determine the spin susceptibility, $\chi$, by the tilted field
method\cite{Fang}, in the context of GaAs HIGFETs\cite{Zhu}. We
briefly review the method, referring to Ref.\cite{Zhu} for details
: For a 2D system in a magnetic field, the spacing between Landau
levels, $\hbar\omega_{c}=e\hbar B_{\bot}/m^{*}$, depends on the
perpendicular field $B_{\bot}$, while the Zeeman energy $\Delta
E_{z}=g^{*}\mu_{B}B_{tot}$, depends on the total magnetic field,
$B_{tot}$. When $|g^{*}\mu_{B}B_{tot}|=i\hbar\omega_{c}$, where
the coincidence index, $i$, is an integer or a
half-integer\cite{footnote}, the Landau levels of the spin-up and
spin-down electrons form a ladder of evenly spaced energy levels.
A schematic diagram of evenly spaced energy levels is shown for
$i=1/2$ as an insert to figure 1.

\begin{figure}
\includegraphics[width=75mm]{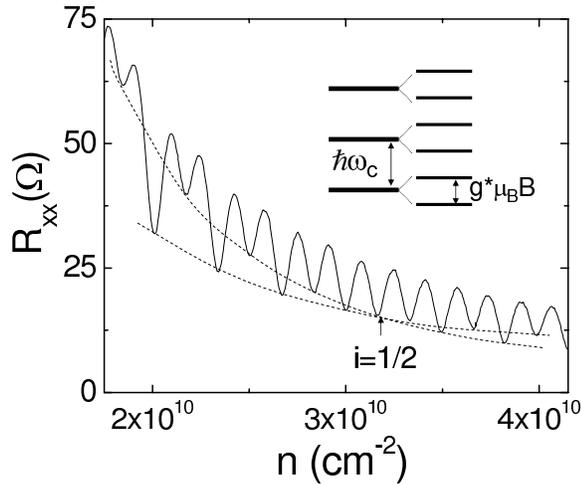}
\caption{\label{fig:SdH} A typical gate sweep trace showing the
coincidence condition for $i=1/2$, at $B_{\bot}=0.07T$ and tilt
angle=$86.5^{o}$. The arrow indicates the density for coincidence
to be $n=3.3\times10^{10}cm^{-2}$. inset: schematic diagram for
the $i=1/2$ coincidence, where spin splittings from the same and
neighboring Landau levels are evenly spaced.}
\end{figure}

The coincidence condition has a clear signature in the oscillation
amplitudes of the magnetoresistance. In a density sweep at fixed
$\theta$ and $B$, and assuming m* and g* not to depend on density,
such evenly spaced energy levels reveal themselves as uniform
oscillation of uniform amplitudes. However, in general, m* and g*
depend on electron density and therefore such equal amplitude
oscillations only appear at a particular density, $n$. Only at
this particular $n$ the value of
$m^{*}g^{*}=2im_{0}B_{\bot}/B_{tot}$. As an example, fig. 1 shows
such a sweep for the i=1/2 coincidence.

\section{\label{sec:level1}Results}

Fig. 2 shows the so determined $n$-dependent values of m*g*. Our
data cover low $r_{s}$ values and, hence, persist into the low
interaction regime, whereas earlier data by Zhu et al.(also shown)
are limited to the high interaction regime($r_{s}\gtrsim3$). Our
data were taken in two different perpendicular fields
$B_{\bot}=0.1T$ and $B_{\bot}=0.07T$, both using the coincidence
index $i=1/2$. The error bars in the y-direction represent the
uncertainties in angle measurements. For comparison, we also show
the low-density data for the $i=1/2$ taken at $B_{\bot}=0.07T$
from Ref.\cite{Zhu}. The small difference between our
$B_{\bot}=0.07T$ data and theirs is most likely due to a slightly
thinner wave function thickness owing to the extra AlAs layer in
our sample(see above).

\begin{figure}
\includegraphics[width=90mm]{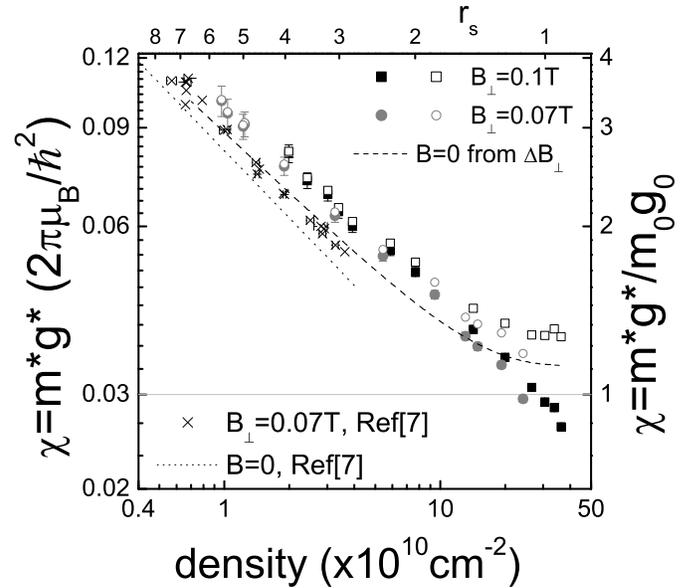}
\caption{\label{loglog} Spin susceptibility $\chi$ versus electron
density. Black squares are measurements for $i=1/2$ in
perpendicular field, $B_{\bot}=0.1T$; grey circles are data for
$i=1/2$ and $B_{\bot}=0.07T$. The error bars in $\chi$ represents
the uncertainties in angle. Errors in density are smaller than the
symbol size. Open symbols represent data of same-shaped filled
symbols after corrected for band structure effects. The
extrapolation of $\chi$ to $B_{tot}\rightarrow 0$ is plotted as a
dashed line(see text). After this correction the right scale
becomes effectively $\chi=m^{*}g^{*}/m_{b}g_{b}$, which is used in
fig. 4. Low density data from ref.\protect\cite{Zhu} are plotted
for comparison.}
\end{figure}

To put the main features of our data in perspective, we first
follow the conventional way of normalizing the $\chi$ data simply
to the literature band edge values of $m_{0}g_{0}$. We provide
this scale on the right side of Fig.2 Clearly the spin
susceptibility decreases monotonically with increasing density
from 3.4 to 0.88 for electron densities from $0.1\times 10^{11}$
to $4\times 10^{11} cm^{-2}$. On this log-log plot, the data from
each $B_{\bot}$ follow a power law behavior with a similar slope
as observed earlier in the low density regime \cite{Zhu}.
Obviously the data drop below $\chi=1$ for high densities, which
is unphysical and contradicts all theoretical predictions.
However, bandstructure effects, which modify the ``bare" mass and
g-factor of the carriers, alter the powerlaw and prevent this from
occurring. While such effects are negligible in the low density
regime, they are appreciable at higher densities and need to be
corrected for. In particular, in GaAs, the band mass $m_{b}$
increases with increasing density, while the magnitude of the
g-factor, $|g_{b}|$, decreases, with the latter being dominant.

\section{\label{sec:level1}Corrections for band structure effects}

\begin{figure}
\includegraphics[width=75mm]{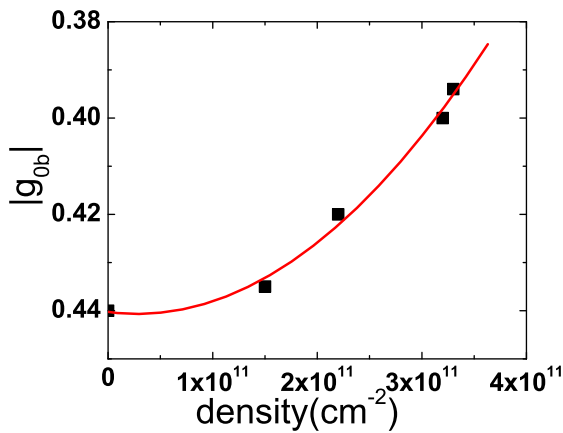}
\caption{\label{g0} Data and fit for coefficient $g_{0b}$ from
literature values\protect\cite{Dober, Dober2, Meisels} (see text
and equation (1)).}
\end{figure}

The density dependence of the electron band mass in a 2DES in GaAs
due to band nonparabolicity has been widely studied and can be
readily calculated \cite{nonparabolicity}. For our density range
$0.1\times 10^{11}$ to $4\times 10^{11} cm^{-2}$, the enhancement
of $m_{b}$ grows from $0.3\%$ to $5\%$. The density dependence of
the $g$-factor is more complex. In a 2DES, this dependence was
first observed by Stein et al. \cite{Stein} using Electron Spin
Resonance (ESR) and was later studied systematically via ESR by
Dobers et al.\cite{Dober}. They deduced an empirical formula
$g_{b}(B, N)=g_{b0}-c(N+1/2)B$, where N is the Landau level index,
$B$ is the magnetic field, and $g_{b0}$ and c are two coefficients
that may change with electron density. The observation was
explained theoretically \cite{Lommer} as a combination of
nonparabolicity of the band structure and an increase of the
penetration of the electron wavefunction into the AlGaAs layer.
Since the $g$-factor of $Al_{x}Ga_{1-x}As$ changes considerably
with Aluminum concentration and has a different sign from $g_{0}$
in GaAs, the magnitude of the $g$-factor shrinks with increasing
penetration of the electron wavefunction. In order to correct our
data, we use the existing measurements\cite{Dober, Dober2,
Meisels} of $g_{b0}$ and c values from samples with electron
densities from $n=1.5\times 10^{11}$ to $n=3.3\times 10^{11}
cm^{-2}$. Adding $g_{b0}=-0.44$ for $n\rightarrow 0$, we
interpolate and achieve the phenomenological dependence,
\begin{equation}
g_{b0}(n)=-0.44-2.75\times 10^{-4}n+4.98\times10^{-5}n^{2},
\end{equation}
where $n$ is in $10^{10}/cm^{-2}$(see fig.3). The coefficient c is
assumed constant, c=0.012, since it hardly varies in
Ref.\cite{Dober}. While Dobers et al. used an Al concentration of
$\sim35\%$, the interface in our sample consists of $100\%$ Al.
From band structure calculations we can estimate that such a
$100\%$ Al concentration at the interface reduces the g-factor
renormalization by $\sim1/2$ as compared to $\sim35\%$ Al. When
combined with a smaller penetration into the AlAs layer, the final
enhancement $m^{*}g^{*}/m_{b}g_{b}$ is affected by less than $4\%$
at our highest $r_{s}$ and has no observable effect for
$r_{s}>1.5$. A similar estimate yields a variation of c by less
than $1\%$ from $Al_{0.35}Ga_{0.65}As/GaAs$ to $AlAs/GaAs$. These
calculations are approximate. However, we deem them sufficient
since their impact on our data resides within the experimental
error bar.

The open symbols in fig. 2 represent all our $\chi$ data corrected
for the density dependence of mass and g-factor. As required, all
data remain above $\chi=1$ and they seem to be levelling off as n
increases. Since we have been working with a small spin
polarization ($i=1/2$) our data can be assumed to be close to the
unpolarized(\textit{i.e.} $B_{tot}\rightarrow 0$) case, which is
the condition used in most theoretical calculations and therefore
of importance for comparisons with experiment. Based on the
$B_{tot}$-dependence of our low-density data we can actually
provide an extrapolation scheme to $B_{tot}\rightarrow 0$, which,
since it is a small correction, should hold over the extended high
density range provided here.

\begin{figure}
\includegraphics[width=90mm]{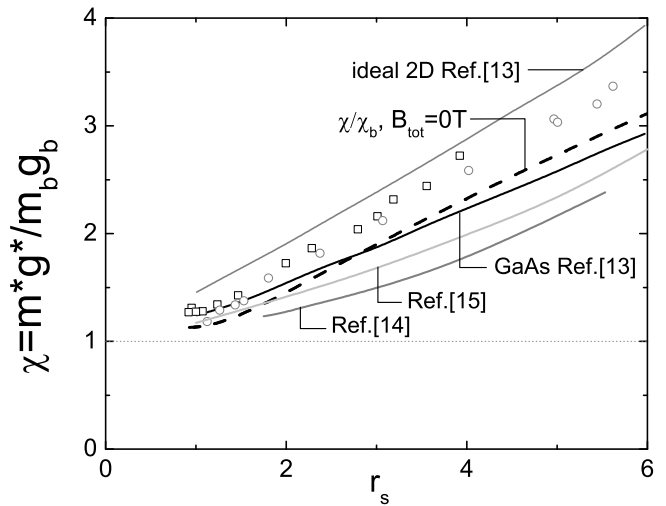}
\caption{\label{rsplot} Normalized spin susceptibility enhancement
$\chi$ versus $r_{s}$. The open symbols are data corrected for
band structure effects due to nonparabolicity. The extrapolation
to $B_{tot}=0$ is shown as a dashed line as in fig.2. $m_{b}$ and
$g_{b}$ denote the bare m and g values after corrected for band
structure effect. The solid curves of different grey scale are
theoretical calculations from ref.\protect\cite{dePalo, Zhang,
Dharma}. The topmost curve is for a zero-thickness 2DES, whereas
all other solid curves are for finite thickness GaAs systems.}
\end{figure}

We observe that the $B_{\bot}=0.1T$ data and the $B_{\bot}=0.07T$
data are slightly offset from each other in the vertical
direction. In particular, at lower densities both follow the
previously observed powerlaw, which had allowed for a simple
extrapolation scheme\cite{Zhu} to $B_{tot}\rightarrow 0$. This
extrapolation, based on different coincidence indices, $i$, can
directly be translated to different $B_{\bot}$ and identical
$i=1/2$ as is the case in the present measurements (see footnote
\cite{footnote2}). The result is a constant multiplicative factor,
which appears as a small downward shift on our log-log plot of
Fig. 2 is shown as a dashed line. While this extrapolation scheme
is derived in the range of the powerlaw dependence between $\chi$
and n of our data, we expect it to apply approximately also in the
range of saturating $\chi$ towards higher densities.

An enhancement of m* and a deduced increase of g* by a high
in-plane magnetic field has been reported by Tutuc et
al.\cite{Tutuc2} According to their observations, the mass
enhancement for our largest in-plane field of $3.9T$, at density
$3.6\times 10^{11}cm^{-2}$, is approximately $8\%$. Extrapolating
their g* enhancement to this density and field, we find it to be
less than $\sim2\%$. Our $\chi$ enhancement from the
$B\rightarrow0$ extrapolation is $\sim9\%$, which is close to the
overall $\sim10\%$ increase deduced from Ref.\cite{Tutuc2}. This
supports that our method, which \textit{extrapolates} the spin
susceptibility to vanishing $B$ field, $\chi_{B\rightarrow
0}=m^{*}g^{*}/m_{0}g_{0}$, is not sensitive to in-plane field
effects on m* and g*.

Fig 4 shows our final results as $\chi$ versus $r_{s}$ , rather
than versus density, for better comparison with theory. The
symbols and lines refer back to fig.2. The data points as well as
the extrapolated curve show the correct limiting behavior of
$\chi\rightarrow 1$ as $r_{s}\rightarrow 0$ as a consequence of
the correction of the data for band structure effects. Without
such correction, the $\chi$ data would have fallen below $\chi=1$,
which would be clearly unphysical.

\section{\label{sec:level1}Comparison with Theories}

Finally, we can compare our data with theory. In the past two
decades, there have been several calculations of the spin
susceptibility $\chi$, but most of them evaluate $\chi$ in the
limit of zero polarization in an ideal 2DES. A figure summarizing
those theoretical results can be found in Ref.\cite{Attaccalite}.
The calculations from different techniques are in rather good
agreement for $r_{s}<3$ but deviate considerably for large
$r_{s}$. In the same regime, measurements of $\chi$ in different
2DESs have also shown considerable discrepancies. A recent report
of dePalo et al, attributes many of these discrepancies to the
finite thickness of the real 2DES and variations of this thickness
between 2DESs from different material systems. For comparison with
our data, extrapolated to $B\rightarrow 0$ (dashed line), we have
included in fig.4 the zero thickness result and the finite
thickness result for GaAs from ref\cite{dePalo}. The calculations
from Zhang and Das Sarma\cite{Zhang} using an improved RPA and
Dharma-wadana\cite{Dharma} using a CHNC method both reside further
below our data.

As can be seen from fig.4, our experimental data and the
theoretical calculations for a realistic 2DES in GaAs show rather
good agreement. For high $r_{s}$ values, this had been pointed out
before\cite{dePalo}. Here we show that after appropriate
correction for band structure effects, one also achieves good
agreement in the intermediate interaction regime ($r_{s}\sim1$)
and obtains the correct trend for $r_{s}\rightarrow0$.

\section{\label{sec:level1}Conclusion}

In conclusion, our data on the spin susceptibility of a 2DES in
GaAs over a very wide of electron densities, covering $r_{s}$
values from strong electron interactions ($r_{s}\sim6$) through
the intermediate interaction regime ($r_{s}\sim1$), and into the
low interaction regime show the correct trends for the limiting
behavior ($r_{s}\rightarrow0$) of this quantity. Only when theory
takes into account the finite thickness of the real 2DES and
experiment is corrected for band structure effects, a rather good
agreement between data and calculation is reached over the whole
range of $r_{s}$ of this study. Our work highlights the fact that
subtleties of the real 2DES need to be taken into account when
comparison with theoretical calculations are being made.

\begin{acknowledgments}
The authors acknowledge Dr. Roland Winkler for discussion. This
work is supported by NSF under DMR-03-52738, by the DOE under
DE-AIO2-04ER46133, and by a grant of the W. M. Keck Foundation.
\end{acknowledgments}


\begin{thebibliography}{24}
\bibitem[{foo({\natexlab{a}})}]{footnote}
\bibinfo{note}{The g-factor does depend on Landau level filling and hence this
  equation is only approximately observed. However, since we are working at
  relatively high densities, low perpendicular field and in high Landau levels,
  the oscillation amplitude of the g-factor is small and we can assume it to be
  constant near the Fermi surface.}
\bibitem[{foo({\natexlab{b}})}]{footnote2}
\bibinfo{note}{Powerlaw fits to our two data sets at low n and
different $B_{\bot}$ yield
$m^{*}g^{*}/m_{0}g_{0}(B_{\bot}=0.1T)=3.40n^{-0.365}$ and
$m^{*}g^{*}/m_{0}g_{0}(B_{\bot}=0.07T)=3.27n^{-0.365}$. Based on
ref.[7], we take the prefactor in the power law to be linear in
total field $B_{tot}$ as $3.40=a+bB_{tot}^{0.1T}$,and
$3.27=a+bB_{tot}^{0.07T}$. Using
$i=1/2=m^{*}g^{*}\mu_{B}B_{tot}/e\hbar B_{\bot}$ to link
$B_{\bot}$ with $B_{tot}$ we arrive at
$B_{\bot}=0.0295(m^{*}g^{*}/m_{0}g_{0})B_{tot}$, yielding
$B_{tot}^{0.1T}/B_{tot}^{0.07T}=1.37$. Solving for
$a=[3.40-(B_{tot}^{0.1T}/B_{tot}^{0.07T})3.27]/
[1-(B_{tot}^{0.1T}/B_{tot}^{0.07T})]=2.92$, and applying the
solutions to $m^{*}g^{*}/m_{0}g_{0}=(a+bB_{tot})n^{-0.365}$ we
arrive at $\chi_{B\rightarrow 0}=2.92n^{-0.365}$ as
$B_{tot}\rightarrow 0$ for low n. This modified prefactor leads to
a vertical shift and is applied to all data to yield
$\chi(B_{tot}\rightarrow 0)$.}
\end{thebibliography}

\end{document}